\documentstyle[epsfig,paspconf]{article}
\def\be{\begin{equation}}
\def\ee{\end{equation}}
\def\bea{\begin{eqnarray}}
\def\eea{\end{eqnarray}}

\def\etal{{\em et al.}            }

\def\mum{\mu {\rm m}}
\def\kms{{\rm \,km\,s}^{-1}}

\def\Ha{${\rm H}\alpha$}

\def\dg{^{\circ}}

\def\Jy{{\rm \,Jy}}
\def\MJy{{\rm \,MJy}}
\def\spose#1{\hbox to 0pt{#1\hss}}
\def\simlt{\mathrel{\spose{\lower 3pt\hbox{$\mathchar"218$}}
     \raise 2.0pt\hbox{$\mathchar"13C$}}}
\def\simgt{\mathrel{\spose{\lower 3pt\hbox{$\mathchar"218$}}
     \raise 2.0pt\hbox{$\mathchar"13E$}}}
\def\({\left(}
\def\){\right)}
\def\[{\left[}
\def\]{\right]}
\def\<{\left\langle}
\def\>{\right\rangle}

\def\AstAst{{\em Astron.\ Astrophys.~}}
\def\AAS{{\em Ast. Ast. Supp.~}}
\def\AJ{{\em Astron.~J.~}}
\def\ApJ{{\em Astrophys.~J.~}}

\def\ApJSupp{{\em Astrophys.~J.\ Supp.~}}

\def\IAU130{in {\em Large Scale Structures of the Universe}, IAU Symposium 130~}
\def\MN{{\em Mon.Not.R.astr.Soc.~}}

\def\edcomment#1{\iffalse\marginpar{\raggedright\sl#1\/}\else\relax\fi}
\reversemarginpar

\begin{document}
\title{The PSCz catalogue}
%\author{Will Saunders$^1$, Seb Oliver$^2$, Oliver Keeble$^2$, Michael Rowan-Robinson$^2$, \\Steve Maddox$^3$, Richard McMahon$^3$, George Efstathiou$^3$, \\Will Sutherland$^4$, Helen Tadros$^4$, Simon White$^5$, Carlos Frenk$^6$}
%\affil{1. Institute for Astronomy, University of Edinburgh \\ 2. Imperial College, University of London \\ 3. Institute of Astronomy, Cambridge University \\ 4. Department of Physics, Oxford University \\ 5. MPI-Astrophysik, Garching \\ 6. Department of Physics, University of Durham}

\author{Will Saunders$^1$, Seb Oliver$^2$, Oliver Keeble$^2$, Michael Rowan-Robinson$^2$} 
\vspace{-5pt}
\affil{1. Institute for Astronomy, University of Edinburgh \\ 2. Imperial College, University of London}
\author{Steve Maddox$^3$, Richard McMahon$^3$, George Efstathiou$^3$}
\vspace{-5pt}
\affil{3. Institute of Astronomy, Cambridge University}
\author{Will Sutherland$^4$, Helen Tadros$^4$, Simon White$^5$, Carlos Frenk$^6$} 
\vspace{-5pt}
\affil{4. Department of Physics, Oxford University \\ 5. MPI-Astrophysik, Garching \\ 6. Department of Physics, University of Durham}

\begin{abstract}
We present the catalogue and redshift data for the PSCz survey of 15,000 IRAS galaxies over 84\% of the sky. The uniformity, completeness and data quality are assessed, and guidelines and caveats for its use are given.
\end{abstract}
\vspace{-10pt}
\section{Introduction}
\vspace{-5pt}
The PSCz survey was started in 1992, when accumulation of IRAS redshifts had made a complete redshift survey of the Point Source Catalog (Beichman \etal 1984, ES) possible. Our goals were to (a) maximise sky coverage, and (b) to obtain the best possible completeness and flux uniformity within well-defined area and redshift ranges. The availability of digitised optical information allowed us to use conservative IRAS selection criteria, and use optical identification as part of the selection process. The sky coverage is essentially limited only by requiring that optical extinction be small enough to allow complete identifications. The PSC was used as starting material, because of its superior sky coverage and treatment of confused and extended sources as compared with the Faint Source Survey (Moshir \etal 1989, FSS). The depth of the survey ($0.6\Jy$) derives from the depth to which the PSC is complete over most of the sky. 
\section{Construction of the Catalogue}
\vspace{-5pt}
The basis for the PSCz was the QIGC (Rowan-Robinson \etal 1990). However, many additions and changes were made to improve the completeness, uniformity and sky coverage.

\vspace{5pt} \parindent=0pt {\bf Sky coverage. \\} The mask includes the coverage gaps, areas flagged as High Source Density at $12$, $25$ or $60\mum$, and the LMC and SMC. We also masked all areas with $I_{100} > 25 \MJy/ster$ or extinction $A_B>2^m$, as
 derived from Rowan-Robinson \etal (1991) and Boulanger and Perault (1988), and including a simple model for dust temperature variations across the galaxy. The overall PSCz coverage is 84\% of the sky. For statistical studies of the IRAS galaxy distribution, where uniformity is more important than sky coverage, we made a `high $|b|$' mask, as above but including all areas with $A_B>1^m$, and leaving 72\% of the sky.

\vspace{5pt} \parindent=0pt {\bf PSC selection criteria.\\} Our aim was to include virtually all galaxies, even at low latitude, purely from their IRAS properties while keeping contamination by Galactic sources to a reasonable level. Our final selection criteria were:

\begin{tabular}{llcr} 
1.& $log(S_{60}/S_{25})$ &$>$& $-0.3$ \\
2.& $log(S_{25}/S_{12})$ &$<$& $1.0$ \\
3.& $log(S_{100}/S_{25})$&$>$& $-0.3$ \\
4.& $log(S_{60}/S_{12})$ &$>$& $0.0$ \\
5.& $log(S_{100}/S_{60})$ &$<$& $0.75$ \\
\end{tabular}

Upper limits were used only where they guaranteed inclusion or exclusion. We made no constraint as to Correlation Coefficient or identification in the PSC. In total, we selected 16885 sources from the PSC.

\vspace{5pt} \parindent=0pt {\bf Extended sources.} \\ All IRAS surveys are bedevilled by the question of how to deal with galaxies which are multiple or extended with respect to the IRAS $60\mum$ beam ($1.5'\times 4'$); in general, there is no consistent
 way to deal with multiple sources in a survey whose beam size is large compared with that of the sources. We chose to preferentially use PSC fluxes, except for sources identified with individual galaxies with extinction-corrected diameter $D_{25c}> 2.25'
$, whose flux is likely to be significantly underestimated in the PSC. To select such galaxies, diameters were taken from the LEDA database (Paturel \etal 1989), extinctions were estimated as for the mask described above, and the resulting corrections to the diameters are taken from Cameron (1990). IRAS fluxes for these galaxies were taken from Rice \etal (1988), or from addscans supplied by IPAC using software supplied by Amos Yahil. The corresponding PSC source was flagged and a new source created using the addscan fluxes. We find that addscan fluxes are systematically 10\% larger than PSC, even for small galaxies; the addscan fluxes quoted in the catalogue have been arbitrarily decreased by 10\%, to bring them statistically into line with the PSC fluxes. 1466 new sources associated with large galaxies entered the catalogue, and 1291 corresponding PSC sources flagged for deletion. Local Group galaxies were excised from the catalogue, and a separate Local Group catalogue made.

\vspace{5pt} \parindent=0pt {\bf Other problem sources.}\\ In all other cases, we opted to use PSC fluxes. We were particularly concerned to avoid introducing any latitude-dependent biases into the catalogue. The confusion flag in the PSC is set very conservatively, and in any case the PSC processing in general deals better with confusion than addscans. Poor Correlation Coefficient generally results from a source being (a) extended and hence dealt with above, or (b) low S/N and/or confused with cirrus, in
 which case the PSC is an unbiased estimator, while addscanning risks noise-dependent biases.

\vspace{5pt} \parindent=0pt {\bf Supplementary sources.}\\ Two separate hours-confirmed detections (HCONS) are required for a source to be included in the PSC. In the 25\% of the sky scanned only twice, the PSC completeness is not guaranteed to $0.6\Jy$ (ES). We
supplemented the catalogue with 1HCON sources with galaxy-like colours
from the Point Source Catalog Reject File (ES), where there is a
corresponding source in the FSS. This revealed many sources where two
individual HCON detections had failed to be merged in the PSC processing,
as well as sources which failed at least one HCON for whatever reason. New
sources were created or merged with existing ones, and the Flux
Correction Factors (ES) recalculated accordingly. We also searched for
additional sources with SES flags 1113, 1122 and 1121. Altogether we found
an additional 490 galaxies, and 143 sources were deleted through
 merging.

\vspace{5pt} \parindent=0pt {\bf Optical identifications.}\\ Optical material for virtually all sources was obtained from COSMOS or APM data, including new APM scans taken of 150 low-latitude POSS-I E plates. In general, we used red plates at $|b|<10\dg$ and blue otherwise. The magnitudes from the digitised sky survey plates are not in general more accurate than $0.5^m$.

At this stage, non-galaxies were weeded out by a combination of optical appearance, IRAS colours and addscan profiles, NVSS maps (Condon \etal 1998), SIMBAD and other literature data, and new $K'$-band imaging data. We found a total of 1527 confirmed non-galaxies. We are left with 15257 confirmed galaxies, and a further 175 where no identification is known but the IRAS properties are not obviously Galactic. The source density amounts to 1460 gals/steradian at high latitudes.

\vspace{5pt} \parindent=0pt {\bf The redshift survey.} \\We maintained a
large database of redshift information from the literature, NED, LEDA and
ZCAT, and also other survey work in progress. Of the 15,000 galaxies in
the sample at the outset, two-thirds either had redshift known or expected
from other surveys.  We were allocated spectroscopic time on the INT (7
weeks), AAT (6 nights), CTIO 1.5m (18 nights) and INAOE 2.1m (2 weeks)
telescopes over a total of 4 years. 4600 redshifts were obtained in this
time, mostly from blended \Ha/${\rm NII}$ lines at low dispersion. The
final external error averages $120 \kms$. 

\section{Reliability, completeness, uniformity, flux accuracy}
\vspace{-5pt}

\vspace{5pt} \parindent=0pt {\bf Reliability.}\\ The major sources of
unreliability in the catalogue are (a) spurious identifications with galaxies
nearby in angular position, and (b) incorrect redshift determination. Selection of targets for spectroscopy used the likelihood methods of Sutherland and Saunders (1992) and the number of incorrect identifications is likely to be very small. For 101 sources, our own redshifts are of marginal quality, and some fraction will be erroneous. Literature redshifts were selected originally on the basis of simple $2'$ proximity (Rowan-Robinson \etal 1990), later cross-correlation used the methods of Sutherland and Saunders, but still depended on quoted positions in the literature. For 143 sources, we have both literature and our own redshifts; 5 are badly discrepant ($\Delta V >1000\kms$), suggesting that about 3\% of the literature redshifts (300 in total) are seriously in error.

\vspace{5pt} \parindent=0pt {\bf Completeness.}\\
1) In the 25\% of sky covered by only 2 HCONs, the PSC incompleteness is estimated as 20\% (differential) and 5\% (cumulative) at $0.6\Jy$ (ES). At $|b|>10\dg$ we estimate that we have recovered $\frac{3}{4}$ of these as supplementary sources (figure 1a). Lower-latitude 2HCON sky (2\% of the area) retains the PSC incompleteness.\\
2) The PSC is confusion limited in the Plane and also affected by noise lagging. However, the mask excludes the worst regions and the source counts shown in figure 1b show no evidence for low-latitude incompleteness to $0.6\Jy$.\\
3) Some galaxies are excluded by our colour criteria. Comparison with the 1.2Jy survey suggests that in total, about 50 galaxies from the PSC have been excluded. \\
4) No attempt was made to systematically obtain redshifts for galaxies with $b_J > 19.5^m$. The distance beyond which this causes incompleteness depends on extinction; we estimate that there is incompleteness for $\log{z} > -1-(0.2(A_B+0.1 A_B^2))$.\\
5) Redshifts are still unknown for 175 galaxies with $b_J < 19.5^m$ (1.1\% of the sample).

\vspace{5pt} \parindent=0pt {\bf Flux accuracy.} \\The error quoted in ES for PSC $60\mum$ fluxes for genuine point sources is 10\%. We are more concerned with any absolute error component, since this will lead to Malmquist-type biases. Lawrence \etal (1999) find an additional absolute average error of $0.059 \pm 0.007 \Jy$. This error leads to Malmquist biases in the source densities which depend on noise level and hence position, especially latitude and 2 vs 3HCON sky. We estimate that the resulting non-uniformity introduced into the catalogue is less than 5\% (differential at the flux limit) and 2\% (cumulative), and this agrees with the source counts (figure 1a). However, note that at some level Malmquist effects will be masked by incompleteness.

\vspace{5pt} \parindent=0pt {\bf Flux uniformity.}\\
1) The absolute calibration of the third 3HCON was revised by a few percent after the release of the PSC (and FSS). The effect of this revision on PSC fluxes would be to change those in the 75\% of the sky covered by 3HCONS by 1-2\%.\\
2)Regions affected by the South Atlantic Anomaly may suffer from calibration problems. However, the source counts for declinations $-40\deg < \delta < 10\deg$, where data is most likely to be affected, shows no evidence for any variation (figure 1a) above the few \% level in source counts.\\
3) Whenever the satellite crossed the Galactic Plane, or other very bright sources, the detectors suffered from hysteresis. Strauss \etal (1990) found that the likely error is typically less than 1\% and always than 2.2\%; This is in agreement with the source counts for identified galaxies in the PSCz as a function of $I_{100}$ (figure 1b).

\vspace{5pt} \parindent=0pt {\bf Summary.}\\ Overall, variations in source
density across the sky due to incompleteness, Malmquist effects and
sensitivity variations are not believed to be greater than a few percent
anywhere at high latitudes for $z<0.1$. Tadros \etal (1999) found an upper limit to the amplitude of large scale,
high latitude harmonic components to the density field of the PSCz of 3\%,
confirming our estimate. At lower latitudes, variations are estimated to
be no greater than 10\% for $ z<0.05$.  

\begin{figure}
\centerline{\epsfig{figure=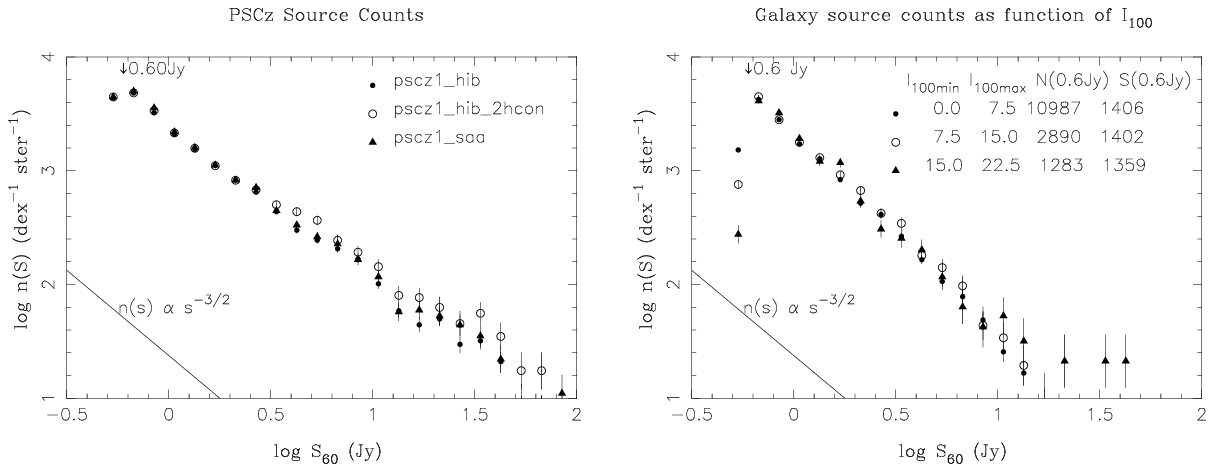,angle=-90}} 
\vspace{5pt} 
Figure 1a. Source counts for all high latitude sky, high
latitude 2HCON sky, and area with potential SAA problems. \\ 
Figure 1b. Source counts for identified galaxies as a function of
$I_{100}$. 
\vspace{-5pt} 
\end{figure}

\end{document}